# Comment on "Spin Transverse Force on Spin Current in an Electric Field"

K.Yu. Bliokh[1,2,3*]

[1]*Institute of Radio Astronomy, 4 Krasnoznamyonnaya st., Kharkov, 61002, Ukraine*
[2]*A.Ya. Usikov Institute of Radiophysics and Electronics, 12 Akademika Proskury st., Kharkov, 61085, Ukraine*
[3]*Department of Physics, Bar-Ilan University, Ramat Gan, 52900, Israel*

In a recent article [1] Shen derived the equations of motion of an electron in an external electromagnetic field from the Dirac equation. He analyzed the operator of the force including a new term of the spin transverse force $\mathbf{F}_f$ (we use notations of [1]), which follows from the non-commutativity of the spin-orbit interaction term. Below we will show that the equations of motion deduced in [1] are incorrect and will adduce semiclassical (i.e. with an accuracy of $\hbar$) expression for the operator of force, which also contains a term analogous to $\mathbf{F}_f$ but twice as large in magnitude. For the sake of simplicity let us consider dynamics of the electron in a scalar potential $V$ (in electric field $\mathbf{E} = -\nabla V/e$) without magnetic field, $\mathbf{A} \equiv \mathbf{B} \equiv 0$. Then, in the semiclassical limit, the operator of force from [1] reads

$$\mathbf{F}^{(Shen)} = -\nabla V + \frac{\hbar}{4mc^2}\mathbf{v}\times\left[(\boldsymbol{\sigma}\nabla)\nabla V - \boldsymbol{\sigma}(\nabla^2 V)\right] + \frac{\hbar}{8m^2c^4}(\boldsymbol{\sigma}\nabla V)(\mathbf{v}\times\nabla V). \qquad (1)$$

It is the last term in Eq. (1) that represents the spin transverse force $\mathbf{F}_f$.

First of all, usual canonical operator of coordinates, $\mathbf{r}$, is used as an operator of coordinates of the electron's center in [1]. However, in the problems where the spin-orbit interaction is essential, the operator of the coordinates of the particle (wave packet) center has the form $\mathit{r} = \mathbf{r} + \hbar\mathcal{A}$ [2–10], where in the case of a non-relativistic electron [9–12] $\mathcal{A} = \mathbf{p}\times\boldsymbol{\sigma}/4m^2c^2$ is $SU(2)$ Berry gauge potential (Berry connection) in the momentum space. Introduction of the operator $\mathit{r}$ leads to non-commutativity of coordinates and provides for covariance of the equations [3–10]. At that, the spin-orbit interaction term in the non-relativistic Hamiltonian of the electron is nothing else than the term $-\hbar\mathcal{A}\dot{\mathbf{p}}$ (dot stands for the full time derivative), which follows from the Berry gauge potential [9–12]. With the covariant coordinate operator, $\mathit{r}$, and gauge potential in the momentum space, $\mathcal{A}$, taken into account, the velocity of the electron takes the form [4,6–10,13]: $\mathbf{v} \equiv \dot{\mathit{r}} = \frac{\mathbf{p}}{m} - \hbar\dot{\mathbf{p}}\times\mathcal{F} = \frac{\mathbf{p}}{m} + \frac{\hbar}{2m^2c^2}\boldsymbol{\sigma}\times\nabla V$. Here $\mathcal{F} = \nabla_\mathbf{p}\times\mathcal{A} - i\mathcal{A}\times\mathcal{A} = -\boldsymbol{\sigma}/2m^2c^2$ is the Berry gauge field (Berry curvature) and the second summand in the equation for $\mathbf{v}$ represents the so-called "anomalous velocity", which is the cause of the intrinsic spin Hall effect [6–10]. This term is exactly twice as large as the similar spin-orbit term that follows from the "naive" non-covariant theory [1] (see [9,10]). The adduced equation shows that the anomalous velocity cannot be interpreted as a gauge potential in the coordinate space, as it is done in [1]. The anomalous velocity contains the Berry gauge field in the momentum space but not a potential in the coordinate one and it is precisely such form of the equation of motion that provides for $SU(2)$ covariance.

---

[*]E-mail: k_bliokh@mail.ru

By using the expression for velocity and standard Hamiltonian $H$ with the spin-orbit interaction term, one can calculate the semiclassical operator of the total force. From $\mathbf{F} = m\dot{\mathbf{v}} = -im\hbar^{-1}[\mathbf{v}, H]$ we have

$$\mathbf{F} = -\nabla V + \frac{\hbar}{2mc^2}\frac{d}{dt}(\boldsymbol{\sigma}\times\nabla V) + \frac{\hbar}{4m^2c^4}(\boldsymbol{\sigma}\nabla V)(\mathbf{v}\times\nabla V), \qquad (2)$$

where $V = V(r)$. The second term in Eq. (2) results from the differentiation of the anomalous velocity and can be represented as $\frac{\hbar}{2mc^2}\left[\left(\boldsymbol{\sigma}\times\frac{\partial\nabla V}{\partial t}\right) + (\mathbf{v}\nabla)(\boldsymbol{\sigma}\times\nabla V)\right]$. This term differs drastically from the second term in Eq. (1), but also vanishes when a homogeneous steady electric field is applied. The last term in Eq. (2) represents a transverse spin force analogous to the last term of Eq. (1), $\mathbf{F}_f$, but twice as large in magnitude. As it is noted in [1], this force is of a purely quantum nature, it comes from the non-commutativity of $\mathcal{A}$ in the spin-orbit term in the Hamiltonian and $\mathcal{F}$ in the anomalous velocity. The transverse spin force cannot be obtained if one substitutes operator $\boldsymbol{\sigma}$ for the classical spin vector in the corresponding semiclassical equations, as it was done in [8,10,13]. This force has the same, first order of smallness in $\hbar$, however it is of the second order in the electric field strength and one should estimate applicability of the semiclassical approximation under consideration.

To conclude, by using correct definition of the operator of the electron's center, we derived semiclassical non-relativistic expression for the force of the electric field action with spin terms taken into account. The obtained expression (2) substantially differs from the Eq. (1) derived in [1]. Thus, the results of [1] require detailed revision, particularly in the case when an external magnetic field is applied and the situation becomes much more complicated [10]. Nevertheless, the basic result of the paper [1] regarding the presence of an additional spin transverse force in the external electric field remains qualitatively true: only numerical factor changes in the equation for $\mathbf{F}_f$.

The work was supported in part by INTAS (Grant No. 03-55-1921).